\documentclass{article}
\usepackage{amsfonts}
\usepackage{epsfig}
\begin{document}

\title{Stochastic and Discrete Time Models of Long-Range Turbulent Transport in
the Scrape-Off Layer } \vspace{1cm}

\author{ {D. Volchenkov} \footnote{St.-Petersburg State University, Research Institute of Physics, St.-Petersburg (Russia).},
{R. Lima} \vspace{0.5cm}\\
{\it  Centre de Physique Theorique, CNRS, Luminy Case 907,}\\
{\it 13288,  Marseille CEDEX 09  France} \\ {\it E-Mail:
dima427@yahoo.com, lima@cpt.univ-mrs.fr}}
\date{\today}

\maketitle

\begin{abstract}
Two dimensional stochastic time model of scrape-off layer (SOL)
turbulent transport is studied. Instability arisen in the system
with respect to the stochastic perturbations of both either
density or vorticity reveals itself in the strong outward bursts
of particle density propagating ballistically across the SOL. The
stability and possible stabilization of the cross-field  turbulent
system depend very much upon the reciprocal correlation time
between density and vorticity fluctuations.

Pdf of the particle flux for the large magnitudes of flux events
can be modelled with a simple discrete time toy model of random
walks concluding at a boundary. The spectra of wandering times
feature the pdf of particle flux in the model and qualitatively
reproduce the experimental statistics of transport events.
\end{abstract}

\vspace{0.5cm}

\leftline{\textbf{ PACS codes: } 05.40.+b, 52.25.Fi, 52.25.Xz}

\vspace{0.5cm}

\leftline{\textbf{ Keywords: } Anomalous transport,
Scrape-Off-Layer, Turbulence stabilization, Stochastic dynamics}

\large

\newpage

\section{Introduction}
\label{Sec1}
\vspace*{-0.5pt}
\noindent

Turbulence stabilization in plasma close to the wall blanket of
the ITER divertor is the important technical problem determining
the performance of the next step device. Long range transport in
the scrape-off layer (SOL) provokes the plasma-wall interactions
in areas that are not designed for this purpose. Evidence of the
strong outward bursts of particle density propagating
ballistically with rather high velocities far  beyond the
$e$-folding length in the SOL  has been observed recently in
several experiments (see \cite{R}-\cite{ADGL}) and in the
numerical simulations \cite{Ghendrih}. These events do not appear
to fit into the standard view of diffusive transport: the
probability distribution function (pdf) of the particle flux
departs from the Gaussian distribution forming a long tail which
dominates at high positive flux of particles \cite{Ghendrih}.

Theoretical investigations of the reported phenomena  remain an
important task. In the present paper, we consider a variety of two
dimensional fluid models based on the interchange instability in
plasma studied in \cite{N}-\cite{G} and discussed recently in
\cite{Ghendrih} exerted to the Gaussian distributed external
random forces to get an insight into the properties of turbulent
transport in the cross-field system.

Neglecting for the dissipation processes in plasma  under the
constant temperatures $T_{e}\gg T_{i}$, this problem is reduced to
the interactions between the normalized particle density field
$n(x,y,t)$ and the normalized vorticity field $w(x,y,t)$ related
to the electric potential field $\phi(x,y,t)$,
\begin{equation}\label{symm}
\partial_t n = [n,\phi], \quad \partial_t w =[w,\phi]-g \partial_y \log n, \quad
w=\Delta_{\bot} \phi, \quad g\geq 0,
\end{equation}
defined in the 2D plane transversal to $\mathbf{e}_z,$ the
direction of axial magnetic field $\mathbf{B}_0$. In (\ref{symm}),
$x$ and $y$ are the normalized radial and poloidal coordinates
respectively. The Poisson's brackets are defined by
$[p,q]=\partial_x p\partial_y q -\partial_x q\partial_y p.$

When $g=0$, the equations (\ref{symm}) describe  the
$2D$-rotations of the density and vorticity gradients around the
cross-field drift $\mathbf{v}=-c/B_0{\ }\nabla\phi \times
\mathbf{e}_z, $ in which $\nabla\equiv (\partial_x,\partial_y)$.
Their laminar solutions (with $w=0$) are given by any spatially
homogeneous electric potential $\phi=\psi_1(t)$ and any stationary
particle density distribution $n=\psi_2(x,y).$ Other
configurations satisfying (\ref{symm}) at $g=0$ are characterized
by the radially symmetric stationary vorticity fields
$w=\partial^2_x \phi(x)$ with the electric potentials invariant
with respect to the Galilean transformation $\phi(x)\to \phi(x) +
x\varphi_1(t)+ \varphi_2(t)$ where the parameters of
transformations $\varphi_{1,2}(t)$ are the arbitrary integrable
functions of time decaying at $t\to -\infty$. The relevant density
configurations $n=\varphi_3\left(x,y-\int_{-\infty}^t
\mathrm{v}_y(x,t'){\ }dt' \right)$ have the form of
profile-preserving waves convected in the poloidal direction by
the poloidal cross field drift $\mathrm{v}_y(x,t).$ The poloidal
component of cross field drift itself remains invariant with
respect to the Galilean transformation $\mathrm{v}_y\to
\mathrm{v}_y+ \varphi_2(t)$, while its radial component
$\mathrm{v}_x=0.$

Configurations that satisfy (\ref{symm}) for $g>0$ have the
Boltzmann density distribution of particles in the poloidal
direction. In particular, those solutions compatible with the
Galilean symmetry discussed above (with $\mathrm{v}_y\neq 0$) are
the solitons (solitary waves) of density convected by the poloidal
electric drift,
\[
n\propto \exp -\frac 1{g\, T(x,y)}\left| y
-\int_{-\infty}^t\mathrm{v}_y(x,y,t')\, dt'\right|,
\]
where $T(x,y)$ is an arbitrary function twice integrable over its
domain. In addition to them, for $g>0,$ Eq.~(\ref{symm}) allows
for the radially homogeneous configurations $\partial_x n=0,$  $\,
\dot{w}=U(y)\, \mathrm{mod} \, 2\pi$ with $\mathrm{v}_y=0$ which
do not fit into the Galilean symmetry, these are the steady waves,
$$
n\propto \exp\, - \frac 1g\, \int^y_0 \, U(y')\, dy'.
$$
The latter solution does not possess a reference angle and can be
considered as an infinitely degenerated state of the system since
the relevant configurations $\{n, \, w\}$ can be made equal at any
number of points by the appropriate choice of $U$:
$U(y_1)=U(y_2)\ldots =U(y_n), $ and $\int^{y_1}_0 \, U(y')\,
dy'=\int^{y_2}_0 \, U(y')\, dy'\ldots =\int^{y_n}_0 \, U(y')\,
dy'.$ For instance, it can be represented by the periodic lattice
potential controlled by the  spokes of high particle density
radiating from the center. With two concurrent symmetries there
can occur either the frustration of one of them or the vanishing
of both with the consequent appearance of a complicated dynamic
picture that is most likely stochastic. The latter case
corresponds to a maximally symmetric motion resulting from the
destruction of unperturbed symmetries \cite{SZ}. In particular,
instability in the system (\ref{symm}) occurs either with respect
to any small perturbation of density or vorticity.

Accounting for the dissipation processes in plasma smears the
picture, so that the small scale fluctuations would acquire
stability. In the present paper, we demonstrate that the small
scales fluctuations can be stable provided there exist the
reciprocal correlations between the stochastic sources of density
and vorticity in the dynamical equations. The large scale
stability of a fluctuation can be characterized by the order
parameter $\xi =\, \left|\,k_y\,\right|\left/\,
\left(k^2_x+k_y^2\right) \right.$  in the momentum space where
$k_x$ and $k_y$ are the radial and poloidal components of momenta
respectively. For the uncorrelated random forces (under the white
noise assumption), a fluctuation with $\xi \,>\,0$ is unstable
with respect to the large scale asymptote in the stochastic
problem.

The accounting for the convection of particles by the random
vortexes introduces a finite reciprocal correlation time
$\tau_c(\,|\mathbf{r}\,-\,\mathbf{r'}|\,)$ between the density and
vorticity random forces. Then there exists the critical scale
$\xi_c,$ in the stochastic model, such that a fluctuation with
$\xi\, <\xi_c$ vanishes with time, but its amplitude grows up
unboundedly with time, for $\xi\, >\,\xi_c$.

Biasing of wall components can locally modify turbulent transport
and is considered to be beneficial if one aims to insulate the
tokamak main chambers from the bursts of density \cite{Ghendrih}.
Indeed, the generation of a uniform electric drift in the poloidal
direction, $\mathrm{v}_y\to \mathrm{v}_y-V,$ would frustrate one
of the symmetries in (\ref{symm}) reestablishing the Galilean
invariance in the system. For instance, those configurations
characterized by the trivial poloidal component of electric drift
$\mathrm{v}_y=0$ would be eradicated. In the present paper, we
have investigated the problem of turbulence stabilization close to
the divertor wall  in the first order of perturbation theory and
shown that there exists a critical value $|\, V_c \,| < \, \infty$
of the poloidal electric drift which would suppress the large
scale instability in the stochastic system with the correlated
statistics of random forces,  $\xi_c\,>\, 0.$ However, for the
uncorrelated random sources in the stochastic problem, $\xi_c\,
=\,0$ and $|\, V_c \,| \to \, \infty$ as $k\, \to\, 0$.

Correlations between the unstable fluctuations of density at
different points are described by the advanced Green's functions
which are trivial for $t\,>\,0$. In particular, these functions
determine the concentration profile of the unstable fluctuations
of density which increases steeply toward the wall. The size of
such fluctuations grows linearly with time. In this case,  the
statistics of the transport events responsible for the long tail
of the flux pdf is featured by the distribution of the
characteristic wandering times of growing blobs convected by the
highly irregular turbulent flow in the close proximity of  the
divertor wall. In our model, we have replaced this complicated
dynamics with the one dimensional (the radial symmetry is implied
in the problem) discrete time random walks. Such a discrete time
model would have another interpretation: the advanced Green's
function is a kernel of an integral equation which relates the
amplitudes of the growing fluctuations apart from the wall with
those on the wall, in the stochastic dynamical problem. Indeed,
this equation is rather complicated and hardly allows for a
rigorous solution. Therefore, being interested in the qualitative
understanding of  statistics of the turbulent transport in the
SOL, we develop a Monte Carlo discrete time simulation procedure
which would help us to evaluate the asymptotic solutions of the
given integral equation.

General approach to the  probability distributions of arrival
times in such a discrete time model has been developed recently in
\cite{FVL}. In general, its statistics can exhibit the
multi-variant asymptotic behavior. Referring the reader to
\cite{FVL} for the details, in the present paper, we have just
shown that the statistics of arrival times for the unstable
fluctuations is either exponential or bounded by the exponentials
(in particular, the latter would be true in the case of the
randomly roaming wall) that is in a qualitative agreement with the
data of numerical simulations and experiments \cite{Ghendrih}.

\section{Stochastic models of  turbulent transport in the cross-filed system.}
\label{Sec2} \vspace*{-0.5pt} \noindent

The stochastic models of cross field turbulent transport used in
the present paper refers to the effectively two-dimensional fluid
model of plasma  based on the interchange instability in the SOL
\cite{N,G} recently discussed in \cite{Ghendrih}. In this model,
one assumes the temperatures of ions and electrons to be constant,
$T_i\ll T_e$. Then the problem is reduced to that of two coupled
fields, the fluctuations of normalized particle density $n(x,y,t)$
and that of vorticity field $w(x,y,t)$, governed by the following
equations
\[
\nabla_t\, w\,=\, u_0\nu\, \Delta_{\bot} w\, - \,\sum_{k\geq 1
}\frac{(-1)^k\, g_k}{k}\,\, \partial_y n^k \, +\, f_w,
\]
\begin{equation}
\label{eq1} \nabla_t \, n \, =\, \nu\,  \Delta_{\bot} n\, +\,  f_n
\end{equation}
written in the polar frame of reference with the normalized radial
$x\, =\, \left.(r - a)\right/\rho_s$ and $y\, =\, \left.
a\theta\right/\rho_s$ poloidal coordinates. Time and space are
normalized respectively to $\Omega_i^{-1},$ the inverse ion
cyclotron frequency, and to $\rho_s$, the hybrid Larmor radius.
The covariant derivative is $\nabla_t\, \equiv\,
\partial_t\, +\, \mathbf{e}_z \cdot \mathbf{v}\, \times \,\nabla,$ in
which $\nabla\, \equiv\, (\partial_x,\, \partial_y)$, and $\Delta_
{\bot}$ is the  Laplace operator defined on the plane transversal
to the axial magnetic field. The effective drive $\propto
\,\partial_y\, \log \left(\left. 1+n\right/ \bar{n}\right)$ risen
in the cross-field system due to the curvature of magnetic lines
is represented  by the series in $\partial_y\, n^k$ with the
coefficients $g_k\,\sim \, \rho_s\left/R \bar{n}^k\right.$ where
$R$ is the major radius of torus and $\bar{n}$ is the mean
normalized particle density. The curvature coefficients  $ g_k$
averaged along the lines of magnetic field are considered to be
constant and small parameters in the problem. The diffusion
coefficients $\nu$ and $u_0\nu $ both are normalized to the Bohm's
value $T_e\,/eB$ and govern the damping in the small scales,
herewith $u_0$ is the dimensionless Prandtl number where the knot
distinguishes its value in the free theory from its effective
value $u$ in the renormalized theory (see Sec.~\ref{Sec4}).

The Gaussian distributed random forces $f_n$ and $f_w$ in
(\ref{eq1}) play the role of stochastic sources in the dynamical
problem maintaining the system  out of equilibrium and
simultaneously modelling the Bohm's boundary conditions at the
sheath which have not been explicitly included in (\ref{eq1}), in
contrast to the original models \cite{Ghendrih}-\cite{G}.
Herewith, the physically important effect of particle escape at
the sheath is replaced with a quenched loss of particles in the
SOL at the points for which $f_n(x,y,\,t)\,<\,0$. Simultaneously,
the particles are supposed to arrive in the SOL in the areas where
$f_n(x, y,\,t)\,>\,0$ modelling the injection of particles from
the divertor core along with the perturbations risen in the system
due to the Langmuir probes \cite{Gunn}-\cite{La1}. For a
simplicity, in the present paper, we assume that the  processes of
gain and loss of particles are  balanced in average therefore $
\left\langle f_n \right\rangle\,=\,0.$ The stochastic source of
particles is used instead of the continuously acting radial
Gaussian shaped  source (localized at $x\,=\,0$) studied in the
numerical simulations \cite{Ghendrih}. Similarly, we impose the
random helicity source $f_w$ exerting onto the vorticity dynamics
in (\ref{eq1}).

Furthermore, the random sources $f_n$ and $f_w$  account for the
\textit{internal} noise risen due to the microscopic degrees of
freedom eliminated from the phenomenological equations
(\ref{eq1}). From the technical point of view, the random forces
help to construct a forthright statistical approach to the
turbulent transport in the SOL. In particular, it allows for the
quantum field theory formulation of the stochastic dynamical
problem (\ref{eq1}) (based on the Martin-Siggia-Rose (MSR)
formalism \cite{MSR}) that gives a key for the use of advanced
analytical methods of modern critical phenomena theory \cite{Ma}.

The Gaussian statistics of random forces in (\ref{eq1}) is
determined by their covariances,
\[
D_{nn}(\mathbf{r}-\mathbf{r'},t-t')\equiv\left\langle
f_n(\mathbf{r},t)f_n(\mathbf{r'},t')\right\rangle, \quad
D_{ww}(\mathbf{r}-\mathbf{r'},t-t')\equiv\left\langle
f_w(\mathbf{r},t)f_w(\mathbf{r'},t')\right\rangle, \quad
\mathbf{r}\equiv (x,y),
\]
describing the detailed microscopic properties of the stochastic
dynamical system. In the present paper, we primarily discuss the
large scale asymptotic behavior of the response functions
$\left\langle\, \left.\delta n (\mathbf{r},t)\right/ \delta f_n
(\mathbf{0},0)\, \right\rangle$ and $\left\langle\, \left.\delta
n(\mathbf{r},t)\right/\delta f_w(\mathbf{0},0)\, \right\rangle$
quantifying the reaction of system onto the external perturbation
and corresponding to the $\mathbf{r}$-distributions of particle
density fluctuations expected at time $t\,>\, 0$ in a response to
the external disturbances of density and vorticity occurring at
the origin at time $t\,=\,0.$ The high order response functions
are related to the analogous multipoint distribution functions
$F_n\,(\mathbf{r}_1,t_1,\,\ldots
\,\mathbf{r}_n,t_n;\,\mathbf{r'}_1,t'_1\,\ldots\,
\mathbf{r'}_n,t_n)$ as
\[
\left\langle \,\frac{\delta^n\,\left[
{n}(\mathbf{r}_1,t_1)\,\ldots\,
{n}(\mathbf{r}_n,t_n)\right]}{\delta{\!}f_n\,(\mathbf{r'}_1,t_1)\,\ldots
\,f_n(\mathbf{r'}_n,t_n)}\,\right\rangle\,
=\,\sum_{\mathrm{permutations}}
F_n\,\left(\mathbf{r}_1,t_1\,\ldots\,
\mathbf{r}_n,t_n;\,\mathbf{r'}_1,t'_1\,\ldots
\,\mathbf{r'}_n,t'_n\right)
\]
with summation over all $n!$ permutations of their arguments
$\mathbf{r}_1,t_1\, \ldots \, \mathbf{r}_n,t_n$.

We consider a variety of microscopic models for the random forces
$f_n$ and $f_w$ in the stochastic problem (\ref{eq1}). Under the
statistically simplest "white noise" assumption, these random
forces are uncorrelated in space and time,
\begin{equation}\label{eq2}
 D_{nn}\,(\mathbf{r}-\mathbf{r'},\,t-t')\,=\,\Gamma_n\,\delta(\mathbf{r'}-\mathbf{r})\,\delta
 (t-t'), \quad
D_{ww}\,(\mathbf{r}-\mathbf{r'},\,t-t')\,=\,\Gamma_w\,\delta(\mathbf{r'}-\mathbf{r})\,\delta
 (t-t'),
\end{equation}
in which $\Gamma_n$ and $\Gamma_w$ are the related Onsager
coefficients.

Recent studies reported on the statistics of transport events in
the cross-field systems \cite{Ghendrih},\cite{PhysPl} pointed out
the virtual importance of correlations existing between density
and vorticity fluctuations in the dynamical problem. In
particular, this effect is referred to the formation of large
density blobs of particles close to the divertor walls by
attracting particles via the cross field flow, the latter being
the larger for strong blobs with strong potential gradients
\cite{Ghendrih}. Indeed, in the physically realistic models of
turbulent transport in the SOL, it seems natural to assume that
the random perturbations enter into the system in a correlated
way. To be specific, let us suppose that there exists a finite
reciprocal correlation time $\tau_c\,\left( |\,\mathbf{r'}
-\mathbf{r}\,|\right)\,>\,0$ between the random sources
$f_w(\mathbf{r},\tau_c)$ and $f_n(\mathbf{r'},0)$ in the
stochastic problem (\ref{eq1}). For a simplicity, we suppose that
the relevant relaxation dynamics is given by the Langevin
equation,
\begin{equation}
\label{L} \frac{\partial f_n}{\partial t}\, =\,
\frac{f_n}{\tau_c}\,+\, \frac {f_w}{\sqrt{\beta}},
\end{equation}
in which $\beta\,\simeq\, \left\langle\,
f^2_w\,\right\rangle\,>\,0$. In the momentum representation, the
non-local covariance operator $\tau^{-1}_c$ can be specified by
the pseudo-differential operator with the kernel
\begin{equation}
\label{kernel}
 \tau^{-1}_c(\,k\,)\,=\,\lambda \nu\,  k^{2-2\gamma}, \quad 0\,<\,2\gamma\,<\,1,
\end{equation}
which specifies the characteristic viscoelastic interactions
between the "fast" modes of density and vorticity fluctuations.
The coupling constant $\lambda\,>\,0$ naturally establishes the
time scale separation between "fast" and "slow" modes. In the case
of $2\gamma\,\ll\, 1,$ the Langevin equation (\ref{L}) with the
kernel (\ref{kernel}) reproduces the asymptotical dispersion
relation typical for the Langmuir waves travelling in plasma,
$\omega \,\sim\, k^{2-\eta_{*}}$ as $k\, \to \, 0$  with
$\eta_{*}\,\simeq \,0.0804$ (in three dimensional space)
\cite{iaw}. Alternatively, for the exponents $2\gamma \,\to\, 1,$
it corresponds to the ion-acoustic waves travelling in the
collisionless plasma with the velocity $\lambda \nu\,\sim
\,\sqrt{T_e\,/\,M}$ where $M$ is the ion mass. Intermediate values
of $\gamma$ correspond to the various types of interactions
between these two types of plasma waves described by the Zaharov's
equations \cite{Zaharov}.

The relaxation dynamics (\ref{L} - \ref{kernel}) establishes the
relation between the covariances of random sources in (\ref{eq1}),
\begin{equation}
\label{solL} D_{nn}(\mathbf{r},t)\,=\,\frac 1{4\pi\beta}\, \int
d\mathbf{r'}dt' \,\int _{0}^{\infty }d\rho\, \frac{J_0 \left( \rho
r' \right)\,\exp\left(-\lambda \nu\,\rho^{2-2\gamma}\, t'\right)
}{\lambda\,
\nu\,\rho^{1-2\gamma}}\,D_{ww}(\mathbf{r'}-\mathbf{r},t'-t) ,\quad
r\,\equiv\, |\,\mathbf{r}\,|,
\end{equation}
where $J_0$ is the Bessel function of the first kind. In the
present paper, we choose the covariance of random vorticity
source,
\[ \left\langle
f_w(\mathbf{r},t)f_w(\mathbf{r'},t') \right\rangle\,= \,\int
\frac{d\omega}{2\pi}\, \int \frac{d\mathbf{k}}{(2\pi)^2}
\,D_{ww}(\omega, k)\,\exp\left[ -i\omega\,
(t-t')+i\mathbf{k}\,(\mathbf{r}-\mathbf{r'})\right], \quad
k\,\equiv\, |\,\mathbf{k}\,|,
\]
in the form of white noise (\ref{eq2}), in which the relevant
Onsager coefficient $\Gamma_w$ is found from the following
physical reasons. Namely, the instantaneous spectral balance of
particle flux,
\begin{equation}
\label{spectral} W(\,k\,) \,=\, \frac 12\, \int
\frac{d\omega}{2\pi}\, \left\langle\, f_n(
\mathbf{k},\omega)\,f_n( \mathbf{-k},\omega)\,\right\rangle,
\end{equation}
derived from (\ref{solL}) should be independent from the
reciprocal correlation time $\tau_c(\,k\,)$ at any $k$ that is
true provided $D_{ww}\,(\omega,\, k)\, \propto\, \lambda\,
k^{-2\gamma}$. Furthermore, the Onsager coefficient $\Gamma_w$ has
to fit into the appropriate physical dimension which is assembled
from the relevant dimensional parameters, $u_0\nu$ and $k$.
Collecting these factors, one obtains the ansatz
$D_{ww}\,\propto\, \lambda \, u_0^3\nu^3 \,k^{6-d-2\gamma},$ in
which $d=2$ is the dimension of space. The power law model for the
covariance of random helicity force $\propto k^{6-d-2\gamma}$ does
not meet the white noise assumption since
$\delta(\mathbf{r}-\mathbf{r'})\,\sim\, k^0$ and therefore calls
for another control parameter $2\varepsilon\,>\,0.$ Eventually, in
the present paper, we use the model
\begin{equation}
\label{Dw3} D_{ww}(k,\,\omega)\, =\, \Gamma_w\,
k^{6-d-2\varepsilon-2\gamma}, \quad \Gamma_w\,\propto \,\lambda\,
u_0^3\nu^3, \quad d\,=\,2,
\end{equation}
with the actual value of regularization parameter
$2\varepsilon\,=\,4,$ for $d\,=\,2.$

Let us note that the ansatz (\ref{Dw3}) is enough flexible to
include the various particular models of particle pump into the
SOL. For instance, the alternative to the white noise assumption
spatially uniform particle pump for which the covariance
\[
D_{ww}\, \simeq \, \delta(\,\mathbf{k}\,)\,=\, \lim_{\xi\to 0}\,
\int d\mathbf{x}\,\left(\frac{x}{\rho_s}\right)^{-\xi}\, e^{i
\mathbf{kx}}=k^{-d}\, \frac{ \Gamma(d/2)}{2\pi^{d/2}}\,
\lim_{\xi\to 0}\left(\xi\, k\,\rho_s\right),
\]
in the large scales, can be represented by the ansatz (\ref{Dw3})
with the actual value $2\varepsilon\,=\,3.$

In the rapid-change limit of the stochastic model, $\lambda\,\to\,
\infty$ (i.e., $\tau_c\,\to\, 0$), the covariance (\ref{solL})
turns into
\begin{equation}
\label{I} \left\langle\, f_n( \mathbf{k},\omega)\,f_n(
\mathbf{-k},\omega)\,\right\rangle\,\simeq\,
  \frac{\nu}{\lambda\,\beta }\, k^{2-d-2\varepsilon+2\gamma},
\end{equation}
and recovers the white noise statistics (\ref{eq2}) along the line
$\varepsilon\,=\,\gamma,$ in $d\,=\,2$. Alternatively, in the case
of $\lambda\,\to\, 0$ (that corresponds to $\tau_c\,\to\,
\infty$), the time integration is effectively withdrawn from
(\ref{solL}), so that the resulting configuration relevant to
(\ref{Dw3}) appears to be static $\propto k^{4-d-2\varepsilon}$
and uncorrelated in space (at $d\,=\,2$) for $2\varepsilon\,=\,2$.

The power-law models for the covariances of random forces has been
used in the statistical theory of turbulence \cite{AAN} (see also
the references therein). The models of random walks in random
environment with long-range correlations based on the Langevin
equation (\ref{L}) have been discussed in concern with the problem
of anomalous scaling of a passive scalar advected by the synthetic
compressible turbulent flow \cite{Ant99}, then in \cite{VCB}, for
the purpose of establishing the time scale separation, in the
models of self organized criticality \cite{BTW}-\cite{Bak1}.

\section{Iterative solutions of the stochastic problem and their diagram representation}
\label{Sec3}
\vspace*{-0.5pt}
\noindent

\begin{figure}[ht]
\noindent
\begin{minipage}[b]{.36\linewidth}
\begin{center}
\epsfig{file=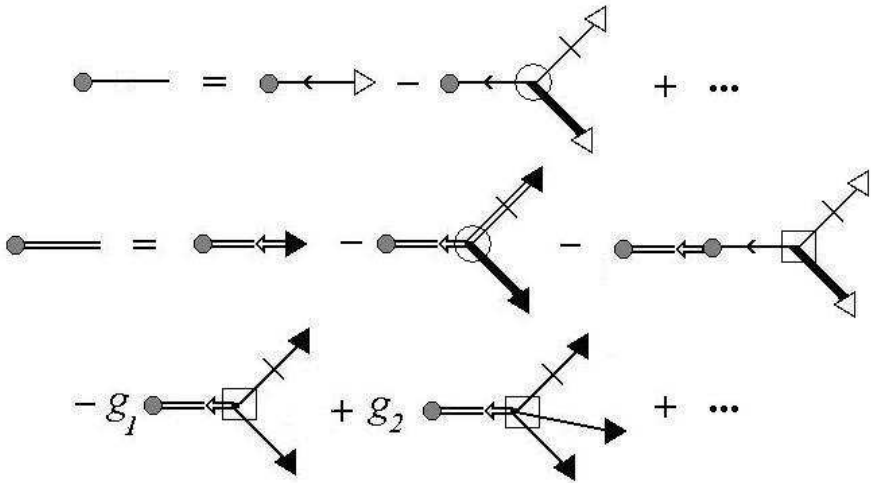, angle= 0,width=12.0cm}
\end{center}
\end{minipage}
\caption{\label{fig1}The diagrammatic representation for
Eq.~(\ref{itsol}).}
\end{figure}

The linearized homogeneous problem, for the fluctuations of
density $n$ and vorticity ${w}$ vanishing at $t\to \infty$,
\begin{equation}
\label{eql}
  \left[\,\partial_t\, -\,\nu \Delta_{\bot}\,\right] \overleftarrow{\Delta}_{n}\,=\,\delta(\,\mathbf{r}\,)\,\delta(\,t\,),\quad
\left[\,\partial_t \,-\,u_0\nu\, \Delta_{\bot}\, \right]
\overleftarrow{\Delta}_{w}(\mathbf{r},\,t)\,+\,g_1\partial_y\,
\overleftarrow{\Delta}_{n}(\mathbf{r},\,t)
\,=\,\delta(\,\mathbf{r}\,)\,\delta(\,t\,),
\end{equation}
is satisfied by the retarded Green's functions,
\[\overleftarrow{\Delta}_n(\mathbf{r},t)
\,=\,\frac{\theta(\,t\,)}{4\pi\,\nu
t}\,\exp\left(\,-\frac{r^2}{4\nu\, t }\,\right),\quad
\overleftarrow{\Delta}_w(\mathbf{r},t)
\,=\,\frac{\theta(\,t\,)}{4\pi\, u\nu \,
t}\,\exp\left(\,-\frac{r^2}{4u\,\nu t
}\,\right)\,+\,\frac{2g_1\,\theta(\,t\,)}{\nu
\,(u+1)}\,\frac{x}{r^2},
\]
fitting into the retarding conditions,
$\overleftarrow{\Delta}_n(\mathbf{r},t)\,=\,\overleftarrow{\Delta}_w(\mathbf{r},t)
\,=\,0$, for  $t\,<\,0$, that express the casualty principle in
the dynamical problem. Nonlinearities in (\ref{eq1}) can then be
taken into account by the perturbation theory,
\[ {n}(\mathbf{r},t)\,=\,\int \,d\mathbf{r'}dt'{\ }
\overleftarrow{\Delta}_n(\mathbf{r}-\mathbf{r'},\,t-t')
\left[\,f_n(\mathbf{r'},t')\,-\,\mathbf{v}(\mathbf{r'},t')\,\times\nabla
{n}(\mathbf{r'},t')\,\right],
\]
\[
{w}(\mathbf{r},t)=\int  d\mathbf{r'}dt'\,
\overleftarrow{\Delta}_w(\mathbf{r}-\mathbf{r'},t-t')
\left[f_w(\mathbf{r'},t')+\sum_{k\geq
1}\frac{(-1)^k\,g_k}{k}\partial_y {n}^k(\mathbf{r'},t')-
\mathbf{v}(\mathbf{r'},t')\times\nabla {w}(\mathbf{r'},t')\right]
\]
\begin{equation}
\label{itsol} +g_1\,\int d\mathbf{r'}dt'\,
\overleftarrow{\Delta}_w(\mathbf{r}-\mathbf{r'},t-t')\,\int
d\mathbf{r''}dt''\,
\overleftarrow{\Delta}_n(\mathbf{r'}-\mathbf{r''},t'-t'')\,
\partial_y\left[ f_n(\mathbf{r''},t'')-\mathbf{v}(\mathbf{r''},t'')\times\nabla
{n}(\mathbf{r''},t'')\right].
\end{equation}
The solutions (\ref{itsol}) allow for the diagram representation
(see Fig.~\ref{fig1}), where the external line (a tail) stands for
the field ${n}$, the double external line denotes the field ${w}$,
and the bold line represents the magnetic flux $\mathbf{v}$. The
triangles stay for the random force $f_n$, and the filled
triangles represent $f_w$. The retarded Green functions
$\overleftarrow{\Delta}_n$ are marked by the lines with an arrow
which corresponds to the arguments $(\mathbf{r'},t')$ and
$(\mathbf{r}'',t'')$ (the direction of arrows marks the time
ordering of  arguments in the lines). Similarly, the double lines
with an arrow correspond to the retarded Green functions
$\overleftarrow{\Delta}_w$. Slashes correspond to the differential
operator $\nabla$. Circles surrounding vertices representing the
antisymmetric interaction $ \mathbf{v}\times \nabla$, squares
present the vertices proportional to the poloidal gradient
$\partial_y.$

\begin{figure}[ht]
\noindent
\begin{minipage}[b]{.36\linewidth}
\begin{center}
\epsfig{file=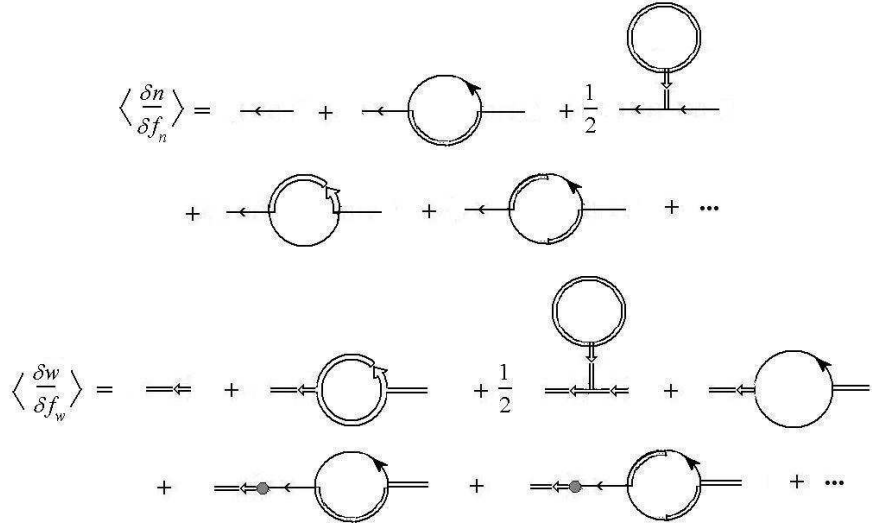, angle= 0,width=12.0cm}
\end{center}
\end{minipage}
\caption{\label{fig2} First diagrams for the simplest response
functions $\left\langle \delta {n}/\delta f_n \right\rangle$ and
$\left\langle \delta {w}/\delta f_w \right\rangle$.}
\end{figure}

All correlation functions of fluctuating fields and functions
expressing the system response for the external perturbations
could be found by the multiplication of trees (\ref{itsol})
displayed on Fig.~\ref{fig1} followed by the averaging over all
possible configurations of random forces $f_n(\,\mathbf{r},t\,)$
and $f_w(\,\mathbf{r},t\,)$. In diagrams, this procedure
corresponds to all possible contractions of lines ended with the
identical triangles. Thereat, the diagrams having an odd number of
external triangles (correspondent to the random forces) give zero
contributions in average. As a result of these contractions, the
following new elements (lines) appear in the diagrams of
perturbation theory:
\[
\Delta_{nn}(\mathbf{r}-\mathbf{r'},t'-t)=\int
d\mathbf{r}_1dt_1\int d\mathbf{r}_2dt_2{\ }
\overrightarrow{\Delta}_n(\mathbf{r}-\mathbf{r}_1,t_1-t){D}_{nn}(\mathbf{r}_1-\mathbf{r}_2,t_1-t_2)
\overleftarrow{\Delta}_n(\mathbf{r}_2-\mathbf{r'},t_2-t'),
\]
\[
\Delta_{ww}(\mathbf{r}-\mathbf{r'},t'-t)=\int
d\mathbf{r}_1dt_1\int d\mathbf{r}_2dt_2{\ }
\overrightarrow{\Delta}_w(\mathbf{r}-\mathbf{r}_1,t_1-t){D}_{ww}(\mathbf{r}_1-\mathbf{r}_2,t_1-t_2)
\overleftarrow{\Delta}_w(\mathbf{r}_2-\mathbf{r'},t_2-t')
\]
\[
+g^2_1\partial^2_y\Delta_{nn}(\mathbf{r}-\mathbf{r'},t'-t),
\]
\[
\Delta_{wn}(\mathbf{r}-\mathbf{r'},t'-t)=g_1\int
d\mathbf{r}_1dt_1{\ }
\overrightarrow{\Delta}_n(\mathbf{r}-\mathbf{r}_1,t_1-t)\partial_y\Delta_{ww}(\mathbf{r}_1-\mathbf{r'},t_1-t'),
\]
\begin{equation}
\label{prop}
\overleftarrow{\Delta}_{wn}(\mathbf{r}-\mathbf{r'},t'-t)=g_1\int
d\mathbf{r}_1dt_1{\ }
\overleftarrow{\Delta}_n(\mathbf{r}-\mathbf{r}_1,t_1-t)\partial_y
\overleftarrow{\Delta}_{w}(\mathbf{r}_1-\mathbf{r'},t_1-t'),
\end{equation}
which are the free propagators of particle density and vorticity
fluctuations, the mixed correlator, and the retarded mixed Green's
function. In diagrams, we present the free propagators
(\ref{prop}) by the correspondent lines without an arrow, and the
retarded mixed Green's functions by the composite directed lines
(see Fig.~\ref{fig2}).

The cross-field drift function $\mathbf{v}(\,\mathbf{r},t\,)$ is
not involved into the linear homogeneous problem (\ref{eql}) and,
therefore, it does not appear in the free propagators
(\ref{prop}), however, it is presented in the nonlinear part of
dynamical equations and therefore appears in the diagrams of
perturbation theory. Due to the simple relation
$\mathbf{w}\,=\,\nabla\,\times\mathbf{v}$, the propagators
containing the field $\mathbf{v}$ are the same as those with
$\mathbf{w}$: $\Delta_{\mathbf{vv}}$, $\Delta_{n\mathbf{v}}$, and
$\overleftarrow{\Delta}_{n\mathbf{v}}$. The bold lines
representing $\mathbf{v}$ in Fig.~\ref{fig1} can be replaced in
the diagrams of perturbation theory with the double lines (which
correspond to the field $\mathbf{w}$) with the additional factor
(in the momentum representation) $-i\,\varepsilon_{zms}\,k_m
/k^2$, where $m,s\, \equiv \, x,y$ and $\varepsilon_{zxy}$ is the
antisymmetric pseudo-tensor, for each $\mathbf{v}$.

In this framework, the exact correlation functions of fields and
the response functions can be found  from the Dyson equations,
\begin{equation}\label{Dyson}
\left\langle\, \frac{\, \delta n\,}{\delta f_n}\,
\right\rangle^{-1}\,=\, \overleftarrow{\Delta}_{n}^{-1}\,-\,
\Sigma_{n}, \quad \left\langle \, \frac{\,\delta {w}\,}{\delta
f_w}\, \right\rangle^{-1}\,=\,
\overleftarrow{\Delta}_{w}^{-1}\,-\, \Sigma_{w},
\end{equation}
where $\Sigma_n$ and $\Sigma_w$ are the infinite diagram series,
in which the first diagrams are shown in Fig.~\ref{fig2}. The
diagram technique introduced in the present section is suitable
for the system preserving the continuous symmetry of (\ref{symm}),
apart from the sheath.

\section{Functional integral formulation of the stochastic problem.
Dimensional analysis, UV divergencies in diagrams and
renormalization}
\label{Sec4}
\vspace*{-0.5pt} \noindent

In the present section, we study the properties of diagram series
resulting from the iterations of the stochastic dynamical
equations with the consequent averaging with respect to all
possible configurations of random forces. The diagrams for some
correlation functions diverge in small scales. The use of
conventional arguments borrowed from the quantum field
renormalization group \cite{ZJ} helps to prove the consequent
subtraction of the logarithmic divergent terms in all orders of
perturbation theory out from the diagrams.

The set of diagrams arisen in the perturbation theory by the
iterations of (\ref{eq1}) is equivalent to the standard Feynman
diagrams of some quantum field theory with the doubled set of
stochastic fields: the fluctuations ${n}$ and ${w}$, the flux
function $\mathbf{v},$ the auxiliary fields $n'$, $w'$
functionally conjugated to the Gaussian distributed random forces
$f_n$ and $f_w$ in (\ref{eq1}), and the Lagrange multiplier
$\mathbf{v}'$ for the binding relation $\mathbf{w}\,=\,\nabla\,
\times \mathbf{v}$. The coincidence of diagrams is a particular
consequence of the general equivalence between the $t$-local
stochastic dynamical problems (in which the interactions contain
no time derivatives) and the  relevant quantum field theories
\cite{Dominicis} with the action functional $\mathcal{S}$ found in
accordance to the MSR formalism \cite{MSR}. Statistical averages
with respect to all admissible configurations of random forces  in
a stochastic dynamical problem can be identified with the
functional averages with the weight $\exp \mathcal{S}.$ In
particular, for the stochastic problem (\ref{eq1}), the generating
functional of the Green functions, $\mathcal{G}(A_\Phi),$ with the
arbitrary source fields $A_\Phi(\mathbf{r},t)$ where $\Phi\equiv
\{{n},n',{w},w', \mathbf{v}, \mathbf{v'}\}$ can be represented by
the functional integral
\begin{equation}
\label{G} \mathcal{G}(A_\Phi)=\int \mathcal{D}\Phi{\ } \exp\left[
\mathcal{S}(\Phi)+\int d\mathbf{r}\, dt {\
}A_\Phi(\mathbf{r},t)\Phi(\mathbf{r},t)\right],
\end{equation}
in which
\begin{equation}
\label{S} \mathcal{S}(\Phi)= \frac 12\int d\mathbf{r}\,
d\mathbf{r'}\, dt \, dt'{\ }\left[
n'(\mathbf{r},t){D}_{nn}(\mathbf{r}-\mathbf{r'},t-t')
n'(\mathbf{r'},t') +
w'(\mathbf{r},t){D}_{ww}(\mathbf{r}-\mathbf{r'},t-t')
w'(\mathbf{r'},t')\right]
\end{equation}
\[
-\,\int \, d\mathbf{r}\, dt{\
}n'(\mathbf{r},t)\,\left[\,\partial_t
{n}(\mathbf{r},t)\,+\,\mathbf{v}(\mathbf{r},t)\, \times \nabla
{n}(\mathbf{r},t)\, -\,\nu \Delta_{\bot} {n}(\mathbf{r},t)\,
\right]
\]
\[
-\int d\mathbf{r}\, dt {\ }w'(\mathbf{r},t)\left[\partial_t
{w}(\mathbf{r},t)+\mathbf{v}(\mathbf{r},t) \times \nabla
{w}(\mathbf{r},t)+\sum_{k\geq 1}\frac{(-1)^kg_k}{k}\,\partial_y
{n}^k(\mathbf{r},t)-u_0\nu \Delta_{\bot} {w}(\mathbf{r},t) \right]
\]
\[
+\, \int\, d\mathbf{r}\, dt {\ }\mathbf{v'}(\mathbf{r},t)\left[\,
{w}(\mathbf{r},t)\,-\, \nabla\,\times \mathbf{v}(\mathbf{r},t)\,
\right].
\]
The source functions $A_{n'}(\,\mathbf{r},t\,)$ and
$A_{w'}(\,\mathbf{r},t\,)$ in (\ref{G}) are interpreted as the not
random external forces, so that the Green functions
$\left\langle\,
{w}(\mathbf{r},t)\,w'(\mathbf{r'},t')\,\right\rangle,$
$\left\langle\,
{n}(\mathbf{r},t)\,n'(\mathbf{r'},t')\,\right\rangle,$ and
$\left\langle\, {n}(\mathbf{r},t)\, w'(\mathbf{r'},t')
\,\right\rangle,$ for $t'\,<\,t$, coincide with the response
functions $\left\langle \,\delta {w}\left/\delta
f_w\right.\,\right\rangle$, $\left\langle\, \delta {n}\left/\delta
f_n\right.\,\right\rangle$, and $\left\langle\,
\delta{n}\left/\delta f_w\right.\,\right\rangle$ respectively. All
possible boundary conditions and the damping asymptotic conditions
for the fluctuation fields $ {n}$ and ${w}$ at $t\,\to -\infty$
are included into the functional integration domain in (\ref{G}).

The functional integral formulation (\ref{G}) of the stochastic
dynamical  problem (\ref{eq1}) allows for the use of various
techniques developed in the quantum field theory to study the
long-time large-scale asymptotic behavior of the quantum and
stochastic systems. The integral (\ref{G}) has the standard
Feynman diagram representation which is equivalent to the
iterative solution of (\ref{eq1}) with an exception of the
self-contracted lines
$\mathrm{Tr}\left(\overleftarrow{\Delta}_{\Phi\Phi'}\right)=
\mathrm{Tr}\left(\overleftarrow{\Delta}_{\Phi'\Phi}\right)=0,$
which are proportional to $\propto \theta (t-t')$ in
time-representation and therefore discontinuous at $t=t'$.
Stipulating that $\theta(0)=0$, one can exclude all redundant
graphs from the perturbation theory of (\ref{G}).  Lines and
vertices in the graphs of perturbation theory are defined by the
conventional Feynman rules and correspond to the free propagators
(equivalent to (\ref{prop})) readily calculated from the free
(quadratic) part of functional (\ref{S}) and the nonlinear
interactions of fields respectively. In the actual calculations,
it is convenient to use the propagators (\ref{prop}) in their
momentum-frequency representation,
\[
\overleftarrow{\Delta}_{nn'}(k,\omega)= (-i \omega+\nu k^2)^{-1},
\quad \overleftarrow{\Delta}_{ww'}(k,\omega)= (-i \omega+u_0\nu
k^2)^{-1},\quad
\Delta_{w\mathbf{v'}}(k)=-{\varepsilon_{zms}k_s}/{k^2},
\]
\[
\Delta_{nn}(k,\omega)=
\overrightarrow{\Delta}_{nn'}D_{nn}(k,\omega)\overleftarrow{\Delta}_{n'n},\quad
\Delta_{ww}(k,\omega)= g_1^2k^2_y\Delta_{nn}(k,\omega)+
\overrightarrow{\Delta}_{ww'}D_{ww}(k,\omega)\overleftarrow{\Delta}_{w'w},
\]
\[
\Delta_{nw}(k,\omega)=i
g_1\overleftarrow{\Delta}_{nn'}(k,\omega)k_y{\
}\Delta_{ww}(k,\omega), \quad
\overleftarrow{\Delta}_{nw'}(k,\omega)= i
g_1\overleftarrow{\Delta}_{nn'}(k,\omega)k_y{\ }
\overleftarrow{\Delta}_{ww'}(k,\omega),
\]
\[
\overleftarrow{\Delta}_{nn'}(k,\omega)=
\overrightarrow{\Delta}^*_{nn'}(k,\omega),\quad
\overleftarrow{\Delta}_{ww'}(k,\omega)=\overrightarrow{\Delta}^*_{ww'}(k,\omega),
\quad
\overleftarrow{\Delta}_{nw'}(k,\omega)=\overrightarrow{\Delta}^*_{w'n}(k,\omega).
\]
Propagators  including the field $\mathbf{v}$ coincide with those
of the field ${w}$ up to the multiplicative factor $
-i\,\varepsilon_{zms}\,k_m/k^2$ for each field $\mathbf{v}.$
Propagators of auxiliary fields $ \Delta_{\Phi'\Phi'}
(k,\omega)=0$.

The action functional (\ref{S}) is invariant with respect to the
generalized Galilean transformations of fields in the poloidal
direction,
\begin{equation}\label{Gt}
{v}_y(\,\mathbf{r},t\,)\,\to\,
{v}_y\left(\,\mathbf{r},t\,\right)\,-\,\mathbf{a}(\, t\, ), \quad
{n}(x,y,t) \,\to\, {n}\left(\,x,y\,-\,\mathbf{b}(\,t\,),t\,\right)
\end{equation}
with the parameter of transformations $\mathbf{a}(\,t\,)$ (the
integrable function decaying at $t\,\to -\infty$) and
$\mathbf{b}(\,t\,)\,=\, \int_{-\infty}^t \,\mathbf{a}(\,t'\,)\,
dt'$. Furthermore, any quantity $Q$ in  (\ref{S}) can be
characterized with respect to the independent scale
transformations, in space and time, by its momentum dimension
$d^k_Q$ and the frequency dimension $d^\omega_Q$. In the
"logarithmic" theory (which is free of interactions that is
analogous to the linearized problem (\ref{eql})), these scale
transformations are coupled due to the relation
$\partial_t\sim\nabla^2$ in the dynamical equations.

This allows for the introduction of the relevant total canonical
dimension $d_Q=d^k_Q+2d^\omega_Q$ and the analysis of UV
divergencies arisen in the diagrams of perturbation theory based
on the conventional dimension counting arguments
\cite{ZJ,Collins}. In dynamical models, $d_Q$ plays the same role
as the ordinary (momentum) dimension in the critical static
problems. Let us note that the poloidal gradient term $\propto
\partial_y$ in (\ref{eql}) is responsible for the stationary
contributions into the Green's function as $t>0$, so that  the
above definition of $d_Q$ remains unambiguous. Stipulating  the
natural normalization conventions, $d^k_k=-d^k_\mathbf{r}=1,$ ${\
}d^\omega_k=d^\omega_\mathbf{r}=0,$ ${\ }d^k_\omega=d^k_t=0,$ ${\
}d^\omega_\omega=-d^\omega_t=1$, one can find all relevant
canonical dimensions from the simultaneous momentum and frequency
scaling invariance of all terms in (\ref{S}) (see Tab.~1).

Integrals correspondent to the diagrams of perturbation theory
representing the 1-irreducible Green's functions $G$ diverge at
the large momenta (small scales) if the correspondent
UV-divergence index $\delta_G$ is a nonnegative integer in the
logarithmic theory,
\begin{equation}\label{delta}
  \delta_{G}=d+2-\left(N_\Phi
  d_\Phi+N_{\Phi'}d_{\Phi'}\right) \geq 0,
\end{equation}
where $d$ is the dimension of space, $d_{\Phi,\Phi'}$ are the
total canonical dimensions of fields $\Phi$ and $\Phi'$, and
$N_{\Phi,\Phi'}$ are the numbers of relevant functional arguments
in $G$. As a consequence of the casualty principle, in the
dynamical models of  MSR-type, all 1-irreducible Green's functions
without the auxiliary fields $\Phi'$ vanish being proportional to
$\theta(0)$ and therefore do not require counterterms \cite{AAN}.
Furthermore, the dimensional parameters  and external momenta
occurring as the overall factors in graphs also reduce their
degrees of divergence (\ref{delta}).

In spite of the following 1-irreducible Green functions could be
superficially divergent at large momenta: $\langle n' {n}
\mathbf{v} \rangle$, $\langle n'\mathbf{v}\rangle$, $\langle n'
\mathbf{v}\mathbf{v} \rangle$, $\langle n'{n}\ldots {n}\rangle$
(with an arbitrary number of fields ${n}$ ), the only Galilean
invariant Green's function admissible in the theory (\ref{S})
which actually diverges at large momenta is $\langle n' {n}
\rangle$. The inclusion of the relevant counterterm subtracting
their superficially divergent contribution is reproduced by the
multiplicative renormalization of the Prandtl number
$$ u_0=uZ_u$$
where $u_0$ and $u$ are the bare and renormalized values of
Prandtl's number. In principle, the relevant renormalization
constant $Z_u$ can be calculated implicitly from the graphs of
perturbation theory up to a finite part of the relevant
counterterm.

However, the standard approach of the critical phenomena theory is
useless for the determining of the large scale asymptotes in the
problem in question, since the severe instability frustrates the
critical behavior in the system preventing its approaching to the
formal asymptote predicted by the conventional renormalization
group method. As a consequence, the critical dimensions of fields
and parameters which can be computed from the renormalization
procedure would have just a formal meaning.

\section{Large scale instability of iterative solutions}
\label{Sec5} \vspace*{-0.5pt} \noindent

The iterative solutions for the stochastic problem (\ref{eq1})
constructed in Sec.~\ref{Sec3} would be asymptotically  stable in
the large scales provided all small perturbations of both density
and vorticity damp out with time. In particular, the exact
response functions found from the Dyson equations (\ref{Dyson})
should have poles located in the lower half-plane of the frequency
space.

\begin{figure}[ht]
\noindent
\begin{minipage}[b]{.36\linewidth}
\begin{center}
\epsfig{file=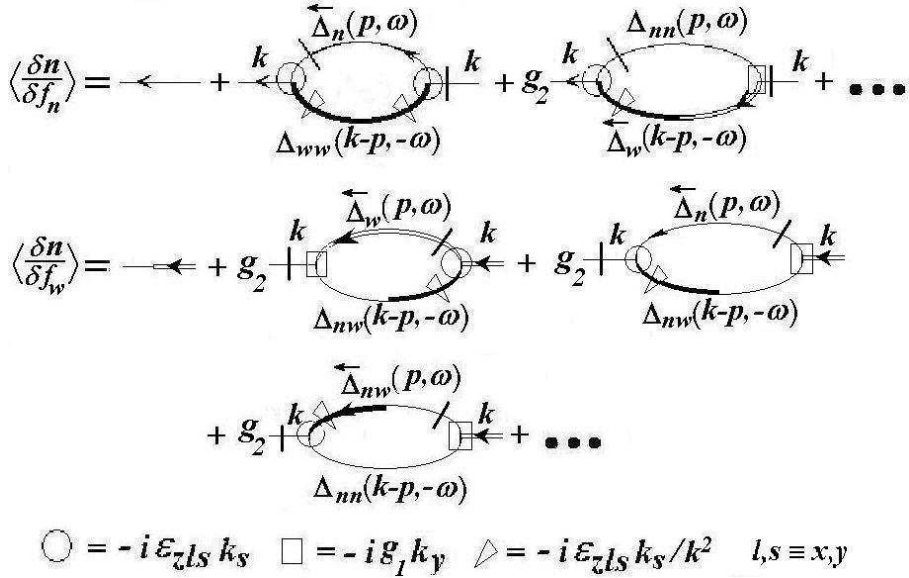, angle= 0,width=12.0cm}
\end{center}
\end{minipage}
\caption{\label{fig3} The simplest 1-irreducible diagrams
contributing into the self-energy corrections $\Sigma_{nn} $ and
$\Sigma_{nw}$ for the response functions $\left\langle \delta
{n}/\delta f_n \right\rangle$ and $\left\langle \delta {n}/\delta
f_w \right\rangle$. The field indices denote the type of
propagators and simultaneously the type of vertexes. The slashes
mark the positions of derivatives $\nabla$, and the skewed
triangles denotes the inverse operator $\mathrm{curl}^{-1}$.}
\end{figure}

The stability of  free response function
$\left.\left\langle\,\delta n\right/\delta f_n\,\right\rangle\,=\,
(-i\omega\, +\,\nu\, k^2)^{-1}$ which effectively corresponds to
the linearized problem (\ref{eql}) is ensured in the large scales
by the correct sign of the dissipation term $\nu k^2\,>\,0$. In a
"proper" perturbation theory, apart from a crossover region, the
stability of exact response functions is also secured by the
dissipation term which dominates over the dispersion relation in
the large scales,
\begin{equation}
\label{resp} \omega(\,k\,) \, =\, -i\,\nu k^2\,+ \, i\,
\Sigma_{nn}(\,k,\omega\,)
\end{equation}
where the self-energy operator $\Sigma_{nn}(\,k,\omega\,)$ is the
infinite series of all relevant $1-$irreducible diagrams of
perturbation theory. However, in the perturbation theory discussed
in Sec.~\ref{Sec3}, the leading contribution into the self-energy
operator is $\Sigma_{nn}\,\simeq\, O(\,k_y\,)$ that could  lift up
the pole of the response function into the upper half-plane of the
complex $\omega-$plane rising the instability in the system as
$k\, \to\, 0$. Such an anomalously strong contribution comes from
the diagrams which simultaneously include both the antisymmetric
interaction vertex $\simeq
i\,\mathrm{e}_z\cdot\,\varepsilon_{ijz}\,\mathrm{v}_ik_j$ together
with the poloidal gradient $\simeq i\, g_k\, k_y $ and the free
propagator of particle density $\Delta_{nn}(\,k,\omega\,)$. Such
diagrams appear in all orders of perturbation theory for the
response functions $\left.\left\langle\,\delta {n}\right/\delta
f_n\,\right\rangle$ and $\left.\left\langle\,\delta
{n}\right/\delta f_w\,\right\rangle$ indicating that the
instability could arise due to the random fluctuations of both
density and vorticity. The first order 1-irreducible diagrams for
them are displayed in Fig.~\ref{fig3}. The small parameters $g_n$
with $n\,>\,2$ appear in the forthcoming orders of perturbation
theory. In particular, the last diagrams in both series shown in
Fig.~\ref{fig3} generate the "anomalous" contributions in the
large scales.

To be certain, let us consider the second diagram in the series
for $\left.\left\langle\,\delta {n}\right/\delta
f_n\,\right\rangle$ which corresponds to the following analytical
expression:
\begin{equation}
\label{int} \left.\Sigma_{nn}\right|_{\mathrm{1-loop}}\,\simeq
\,g_2 k_y \,\int
\frac{d\mathbf{p}}{(2\pi)^2}\,\int\frac{d\omega}{2\pi}\,
\int\frac{d\omega'}{2\pi}
\,\,\frac{k_xp_x+k_yp_y-p_x^2-p^2_y}{(\mathbf{k-p})^2}\,\,
  \Delta_{nn}(\mathbf{p},\omega')\,\,\overleftarrow{\Delta}_w
  (\mathbf{k}-\mathbf{p},\omega-\omega').
\end{equation}
Being interested in the $O(\,k_y\,)$-contribution into
(\ref{int}), one can neglect the $k-$dependence in the integrand.
The analytical properties of  this contribution depends very much
upon the certain assumption on the covariances of random forces
since it changes the free propagator $\Delta_{nn}$. For instance,
under the white noise assumption (\ref{eq2})
$\Delta_{nn}\,\simeq\, (\,\omega^2\,+\,\nu^2 p^4\,)^{-1},$ the
integral in (\ref{int}) diverges at the small scales (large
momenta) for $\varepsilon\,<\,1$ and diverges at the large scales
(small momenta) for $\varepsilon\,>\,1$. Introducing the relevant
cut-off parameters, one obtains the anomalous contribution
\begin{equation}\label{res1}
\left.\Sigma_{nn}\right|_{\mathrm{1-loop}}\,\simeq\,-\,\,
k_y\,{\frac {\nu\, g_{2} \left( {\Lambda}^{2-2\,\varepsilon}-{m}^{
2-2\,\varepsilon} \right) }{8\pi \, \left( -1+\varepsilon \right)
\left( u+1 \right) }},
\end{equation}
in which $u$ is the renormalized value of  Prandtl's number. In
the preceding section, we have shown that the logarithmic
divergencies risen in diagrams for the response function
$\left.\left\langle\,\delta n\right/\delta f_n\,\right\rangle$ in
the small scales (large momenta) can be eliminated from the
perturbation theory by the appropriate renormalization. The
singularity in (\ref{res1}) arisen at the small momenta  $m\to\,
0$ for $\varepsilon\,>\, 1$ would compensate the smallness of
$g_2,$ so that any density fluctuation with $k_y\,>\,0$ appears to
be unstable.

\begin{figure}[ht]
\noindent
\begin{minipage}[b]{.36\linewidth}
\begin{center}
\epsfig{file=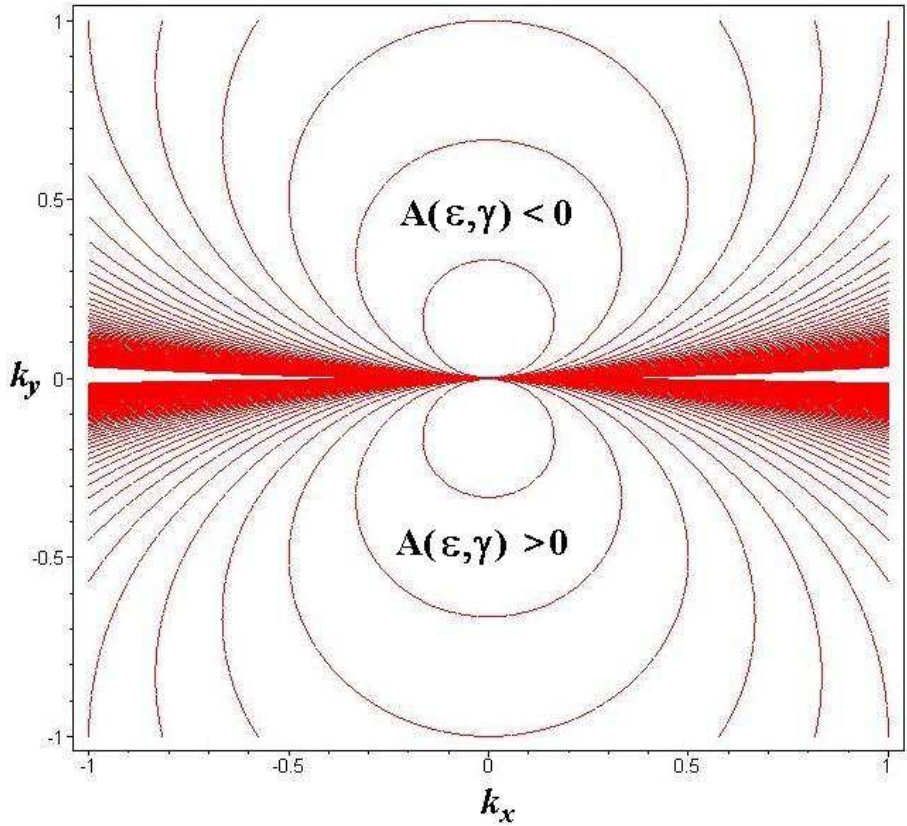, angle= 0,width=12.0cm}
\end{center}
\end{minipage}
\caption{\label{fig4}. The admissibility condition (\ref{resp2})
defines the set of elliptic curves which bound the unstable
segments in the phase space. Density fluctuations with $k_y\,\to\,
0,$ (i.e. extended in the poloidal direction) appear to be stable
in the large scales for any $|\,k_x\,|\,>\,0$. Those fluctuations
characterized by $|\,k_y\,|\,>\,0$  would also be stable in the
large scales provided $ \mathrm{sign}\,
\left(\,k_y\,\right)\,=\,-\,\mathrm{sign}\, \left(\,A(\varepsilon,
\gamma)\,\right)$ for the given values $\varepsilon$ and
$\gamma$.}
\end{figure}

\begin{figure}[ht]
\noindent
\begin{minipage}[b]{.36\linewidth}
\begin{center}
\epsfig{file=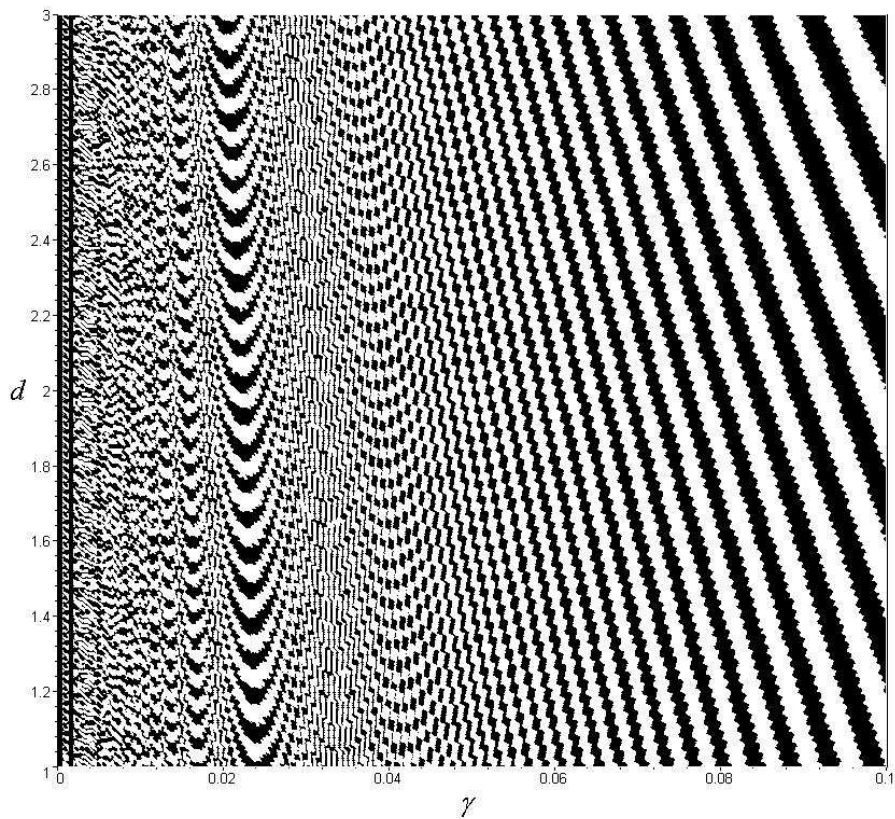, angle= 0,width=12.0cm}
\end{center}
\end{minipage}
\caption{\label{fig5}. The signature of 1-loop order contribution
(\ref{res2}) into $A(\varepsilon, \gamma)$ (black is for $+1,$
white is for $-1$) at different values of $\gamma$ and
$d=6-2\varepsilon-2\gamma$, the space dimension related to the
actual value of regularization parameter $\varepsilon$ under the
statistical assumption (\ref{Dw3}). }
\end{figure}

Accounting for the finite reciprocal correlation time $\tau_c\,>\,
0 $ between the vorticity and density random sources in
(\ref{eq1}) introduces the new dimensional parameter $\lambda>0$
into the particle density propagator, $\Delta_{nn}\,\simeq\,
\left(\,\omega^2\,+\,\nu^2 p^4\,\left(1\,+\,\lambda \,
p^{-2\gamma}\,\right)\,\right)^{-1}.$ Then, the integral
(\ref{int}) can be computed by its analytic continuation for any
momenta excepting for the isolated points, $-1+\varepsilon=\gamma
\, \mathrm{mod}\, 1,$
\begin{equation}
\label{res2}\Sigma_{nn}\, =\, \nu\,
A\,(\,\varepsilon,\gamma\,)\,\, k_y, \quad
 \left. A\,(\,\varepsilon,\gamma\,)
\right|_{\mathrm{1-loop}}
 \,\simeq\, k_y\,\, \frac
18\, \left( {\frac {u}{\lambda}} \right) ^{ (-1+\varepsilon)/
\gamma}\frac{\nu}{{u}}\, \frac{g_{{2}}}{{\gamma}} \csc \left(
{\frac {\pi \, \left( -1+\varepsilon \right) }{\gamma}} \right).
\end{equation}
The dispersion relation (\ref{resp}) determines the region of
asymptotic stability in the phase space of cross-field transport
system. Namely,  a  density fluctuation arisen in the SOL with
some random momenta $(k_x,\,k_y)$ would be asymptotically stable
with respect to the large scales $k\,\to\,0$ if
\begin{equation}
\label{resp2} \frac{\,\,k_y\,\,}{ \,k_x^2\,+\,k_y^2\,}\, < \,
\frac  1{A\,(\,\varepsilon,\, \gamma\,)}
\end{equation}
and be unstable otherwise. In the first order of perturbation
theory, the amplitude factor $A(\varepsilon, \gamma)$ is given by
(\ref{res2}). For different values $A\,(\,\varepsilon,\,
\gamma\,),$ the stability condition (\ref{resp2}) determines the
set of circles (see Fig.~\ref{fig4}) osculating at the origin
which bound the unstable segments of  phase space. One can see
that the density fluctuations with $k_y\,\to\, 0$ (i.e. extended
in the poloidal direction) are asymptotically stable for any
$|\,k_x\,|\,>0$. Density fluctuations characterized by
$|\,k_y\,|\,>0$ would be asymptotically stable in a certain
stochastic model provided $ \mathrm{sign}\,
\left(\,k_y\,\right)\,=\,-\,\,\mathrm{sign}\,
\left(A(\,\varepsilon, \gamma\,)\right),$ for the given values of
$\varepsilon$ and $\gamma$. The signature of the 1-loop order
contribution (\ref{res2}) into $A(\,\varepsilon, \gamma\,)$ is
displayed on the diagram in Fig.~\ref{fig5} (black is for $+1,$
white is for $-1$) at different values of $\gamma$ and
$d=6-2\varepsilon-2\gamma$, the space dimension related to the
actual value of regularization parameter $\varepsilon$ under the
statistical assumption (\ref{Dw3}).

It is important to note that the stability condition (\ref{resp2})
can be formulated as an upper bound for the order parameter $\xi
=\left. \,k_y\,\right/ \,k^2$: $\,\; \xi\,<\,\xi_c$ where $\xi_c
\,=\, A\,(\,\varepsilon,\, \gamma\,)^{-1} $. For the uncorrelated
statistics of random forces, in the stochastic dynamical problem,
 $A(\,\varepsilon, \gamma\,)$ (\ref{res1}) diverges as $k\to 0$ and
therefore $\xi_c\,\to\, 0.$

\section{Turbulence stabilization by the poloidal electric drift}
\label{Sec7}
\vspace*{-0.5pt}
\noindent

To promote the stochastic cross-field turbulent transport system
(\ref{eq1}) from the instability to a stable regime, it seems
natural to frustrate the symmetry which breaks the Galilean
invariancy in (\ref{eq1}). This can be achieved by generating a
constant uniform drift in the poloidal direction
$\mathrm{v}_y\,\to\, \mathrm{v}_y\,-\,V$ (by biasing the limiter
surface, $\phi(x)\, \to \,\phi(x)+\,xV$) that would eradicate
those configurations with the trivial poloidal component of
electric drift $\mathrm{v}_y\,=\, 0.$ In general, the relevant
dispersion equation $\omega\,(\,k,V\,)\,=\,0$ could have many
solutions $V_k$ for $k\, \ll\, 1. $ Herewith, the turbulence
stabilization is achieved for the drifts $V$ from the intervals
$V_{k-1}\,<\,V\,<\, V_k$ for which $\mathrm{Im}(\omega)\,<\, 0.$

To be certain, let us consider the dispersion equation
correspondent to the simplest response function
$\left.\left\langle\, \delta n\right/ \delta f_n \,\right\rangle$,
in the 1-loop order. The leading contribution into the dispersion
equation  is given in the large scale region by the diagram
(\ref{int}). Under the white noise assumption (\ref{eq2}), the
free propagator accounting for the uniform electric drift $V$ is
$\Delta_{nn}\,\simeq\,\left(\,{\omega}^{2}\,+\,{\nu}^{2}{k}^{4}\,+\,{V}^{2}{k_{{y}}}^{2}\,\right)^{-1}$.
Then, for  $|\,k_y\,|\, <\, 1$, the dispersion relation reads as
following,
\begin{equation}\label{disp02}
  \omega(\,k,\,V\,)\,\simeq\,-i\,\nu\,{k}^{2}\,+\,i\,k_{{y}}\,\,\frac{g_{{1}}g_{{2}}}{u+1} \left(\,{\frac {\nu\,
 \left( {m}^{2-2\,\varepsilon} \right) }{8\pi \,
 \left( -1+\varepsilon \right) }}-\,\sqrt{\pi}\,{\frac {\,{V}^{2}
 \left( u+2 \right) \,\Gamma  \left( 1/2+\varepsilon \right)\, \log m }{8\nu\, \left( 1+u \right)\, \Gamma  \left(
2+\varepsilon
 \right) }} \right).
\end{equation}
The latter relation shows that for any finite $m\,>\,0$ there
exists finite $V_c\,<\,\infty$ such that for any $V\,>\, V_c$  one
obtains $\mathrm{Im}(\omega)\,<\, 0$ in (\ref{disp02}), however,
$V_c\, \to\, \infty$ as $m\,\to\,0.$

In contrast to it, in the case of correlated statistics (\ref{L} -
\ref{kernel}, \ref{Dw3}), the dispersion relation is not singular
for $k\,\to\,0$ excepting some particular values of $\varepsilon$
and $\gamma$, and the correspondent stabilizing electric drift, in
the 1-loop order, equals to
\begin{equation}\label{disp03}
  V_c\, =\,\pi \,\nu{u}^{3} \,\,\frac{\, \left( u\lambda \right) ^{1/2{\gamma}} \left(
1+u \right) ^{ 1\,+\,\,(2\varepsilon+1)/2\gamma}\,\,\left(\,
 \left(\, 1+u \,\right) ^{1+\, {\varepsilon}/{\gamma}}-1\, \right)}
{\, \left( 1+u \right) ^{2\,+\,\,(2\varepsilon+1)/2\gamma}+u
\left( 2\,\varepsilon+1 \right) -2\,\gamma\, \left( 1+2\,u
 \right)
\,\,} \,\,\, \frac{\,\sin\,
\left.{\pi\,(\,2\varepsilon\,+\,1\,)}\right/{2\gamma}}{\,\sin\,
{\pi\, \varepsilon}\,/{\gamma}}.
\end{equation}
In the range $0\,<\,\gamma \,<\,1/2$, $\,\, u\,>\,0$, this
expression is singular at the points $\varepsilon\,/\gamma \,
\in\, \mathbb{Z}$.

\section{Qualitative discrete time model of anomalous transport in the SOL }
\label{Sec6}
\vspace*{-0.5pt} \noindent

Large scale instability developed in the cross-field model
(\ref{eq1}) is related to the appearance and unbounded growth of
fluctuations of particle density close to the wall. In accordance
to the fluctuation-dissipation theorem, the fluctuations arisen in
the stochastic dynamical system are related to its dissipative
properties. In particular, the matrix of the exact response
functions $\mathbb{R}(\,\mathbf{k},\omega\,)$ expressing the
perturbations of fields $n$ and $w$ risen due to the random
sources $f_n$ and $f_w$ determines the matrix of exact dynamical
Green's functions $\mathbb{G}(\,\mathbf{k},\omega\,),$
\begin{equation}\label{FDT}
\mathbb{R}\,(\,\mathbf{k},\,\omega\,)\,-
\,\mathbb{R}^{\dagger}\,(\,\mathbf{-k},\,-\omega\,)\,= \,
i\,\omega \, \mathbb{G} \,(\,\mathbf{k},\,\omega\,)
\end{equation}
where $\mathbb{R}^{\dagger}$ is the transposed $\mathbb{R}.$ In
the large scale limit $k\to 0$, we take into account for the
leading contributions into the self-energy operators in the
elements of $\mathbb R$,
\begin{equation}\label{R}
  {R}_{nn}\,\simeq\,
\left( -i\omega\,-\nu\,A_1\, k_{{y}}\,+\nu\,{k}^{2} \right) ^{-1},
\quad R_{nw}\,\simeq\, \left( -i\omega\,-\,\nu\, u \,A_2\,
k_{y}\,+ \nu\,u\,{k}^{2} \right) ^{-1},
\end{equation}
\[
{R}_{wn}\,\simeq\, \left( -i\omega\,+\nu\,{k}^{2} \right) ^{-1},
\quad R_{ww} =\, \left( -i\omega\,+\nu u\,{k}^{2} \right) ^{-1},
\]
in which $A_{1,2}$ are the amplitudes of the anomalous
contributions competing with the dissipation $\propto O(k^2)$ in
the large scales. Fluctuations of particle density arisen in the
model (\ref{eq1}) grow up unboundedly  provided either
$A_1\,k_y\,>\, k^2$ or $A_2\,k_y\,>\, k^2$ for the given values of
$\varepsilon$ and $\gamma.$ The correspondent advanced Green's
functions appear to be analytic in the lower half-plane of the
frequency space,
\begin{equation}\label{Gnn}
  \nu\cdot G^a_{nn}\,(\,\mathbf{k},t\,)=\,\frac{A_1\,k_y\,\, \theta(-t)}{\,k^4\,-\,A_1^2\,
  k_y^2},\quad
\nu\,u\cdot G^a_{nw}\,(\,\mathbf{k},\,t\,)=\,-\frac
{\theta(-t)}{\left( A_2\,k_y\,-\, k^2 \right)} -\frac
{\theta(-t)}{k^2} ,
\end{equation}
being trivial for $t\,>\,0.$

For instance, let us consider the advanced Green's function
$G^a_{nn}$ which relates the density of particles ${\delta\!
n}(\,\mathbf{r},\,t\,)$ in the fluctuations characterized with
$A_1\,k_y\,>\, k^2$ and arisen at the point $\mathbf{r}\in \Omega$
inside the divertor at time $t$ with the particle density
${\delta\! n}(\,\mathbf{r'},\,t'\,)$ of those achieved the
divertor wall at some subsequent moment of time $t'\,>\, t$ at the
point $\mathbf{r'}\in\partial\Omega$:
\begin{equation}
\label{rel01} \left.{\delta\!
n}(\,\mathbf{r'},\,t'\,)\right|_{\partial \Omega} \,=
\int_{\,t\,<\,t'} \, dt \,\, \int_{ \Omega}\,
d\mathbf{r}\,\,{\delta\! n}(\,\mathbf{r},\,t\,)\,G^a_{nn}
(\,\mathbf{r'-r}\,)
\end{equation}
where
\begin{equation}
\label{GR}
G^a_{nn} \left(\mathbf{r'}-\mathbf{r}\right)= \frac 1{2\nu\, A_1
}\,\left[\sin \frac{A_1 (y'-y)}{2}\,{J}_0\left(\frac{A_1
|r'-r|}{2}\right) +\cos\frac{A_1 (y'-y)}{2}\,{H}_0\left( \frac{A_1
|r'-r|}{2}\right)\right],
\end{equation}
in which $\left| r-r'\right|\equiv \sqrt{( x-x')^2+(y-y')^2},$
$J_0$ and $H_0$ are the Bessel and Struve functions respectively.
The integral in the r.h.s. of (\ref{rel01}) is finite for any
$\Omega$ provided $|\,A_1\,|\,>\,1,$ but for the compact $\Omega$
as $|\,A_1\,|\,<\,1$. To be specific, let us consider the circle
$C_R$ of radius $R$ as the relevant domain boundary and suppose
for a simplicity that the density of particles incorporated into
the growing fluctuations inside the domain is independent of time
and maintained at the stationary rate $\delta\! n_0(\,
\mathbf{r})$. Then the $\mathbf{r}-$integral in the r.h.s of
(\ref{rel01}) can be calculated at least numerically and gives the
growth rate $B(R)$ for those density fluctuations,
\begin{equation}
\label{tau}
  {\delta\! n}(\,R,\,\tau\,) \,=\, \tau\,\cdot B(R)\,,
\end{equation}
where $\tau$ is the travelling time of the density blob to achieve
the divertor wall that can be effectively considered as a random
quantity. It is the distribution of such wandering times  that
determines the anomalous transport statistics described by the
flux pdf in our simplified model. The discrete time model we
discuss below is similar to the toy model of systems close to a
threshold of instability studied in \cite{FVL} recently. Despite
its obvious simplicity (the convection of a high density blob of
particles by the turbulent flow of the cross field system is
substituted by the discrete time one-dimensional (in the radial
direction) random walks characterized with some given distribution
function), its exhibits a surprising qualitative similarity to the
actual flux driven anomalous transport events reported in
\cite{Ghendrih}.

We specify the random radial coordinate of a growing fluctuation
by the real number $x\,\in \,[0,\,1].$ Another real number
$R\,\in\,[0,\,1]$ is for the coordinate of wall. The fluctuation
is supposed to be convected by the turbulent flow and grown as
long as $x\,<\,R$ and is destroyed otherwise ($x\,\geq\,R$). We
consider $x$ as a random variable distributed with respect to some
given probability distribution function $\mathbb{P}\{x < u\} =
F(u)$. It is natural to consider the coordinate of wall $R$ as a
fixed number, nevertheless, we discuss here a more general case
when $R$ is also considered as a random variable distributed over
the unit interval with respect to another probability distribution
function (pdf) $\mathbb{P}\,\{R\, <\, u\} = Q(u)$. In general, $F$
and $Q$ are two arbitrary left-continuous increasing functions
satisfying the normalization conditions
$F(\,0\,)\,=\,Q(\,0\,)\,=\,0$,
$F(\,\infty\,)\,=\,Q(\,\infty\,)\,=\,1$.

Given a fixed real number $\eta \,\in\,[0,1]$, we define a
discrete time random process in the following way. At time
$t\,=\,0,$ the variable $x$ is chosen with respect to pdf $F$, and
$R$ is chosen with respect to pdf $Q$. If $x \,<\, R$, the process
continues and goes to time $t\,=\,1$. Otherwise,  provided $x
\,\geq\, R,$ the process is eliminated. At time $t\,\geq\, 1,$ the
following events happen:

\textit{i}) with probability $\eta$, the random variable $x$ is
chosen with pdf $F,$ but the threshold $R$ keeps the value it had
at time $t\,-\,1$. Otherwise,

\textit{ii}) with probability $1\,-\,\eta,$ the random variable
$x$ is chosen with pdf $F$, and  $R$ is chosen with pdf $Q$.

If $x\,\geq\, R,$ the process ends; if $x\,<\,R,$ the process
continues and goes to time $t\,+\,1.$

\medskip

Eventually, at some time step $\tau,$ when the coordinate of the
blob, $x,$ drops "beyond"  $R,$ the process stops, and the integer
value $\tau$ resulted from such a random process limits the
duration of convectional phase. The new blob then arises within
the domain, and the simulation process starts again.

While studying the above model, we are interested in the
distribution of durations of convection phases $P_{\eta}(\tau;\,
F,Q)$  (denoted as $P(\tau)$ in the what following) provided the
probability distributions $F$ and $Q$ are known, and the control
parameter $\eta$ is fixed. The motionless wall corresponds to
$\eta=0.$ Alternatively, the position of wall is randomly changed
at each time step as $\eta =1.$

The proposed model resembles to the coherent-noise models
\cite{NS}-\cite{SN} discussed in connection with a standard
sandpile model \cite{BTW} in self-organized criticality, where the
statistics of avalanche sizes and durations take power law forms.

We introduce the generating function of $P(\,\tau\,)$ such that
\begin{equation}\label{GF}
\hat{P}(s)\,=\,\sum_{\tau=0}^{\infty}\,s^{\tau}\,\,P(\,\tau \,),
\quad P(\tau)\,=\,\left.\frac 1{\,\tau!\,}\,\frac{d^{\tau}\,
\hat{P}(\,s\,)}{ds^{\tau}}\,\right|_{s=0},
\end{equation}
and define the following auxiliary functions
\[
K(n)\,=\,\int_0^{\infty} \,F(u)^n\, dQ(u), \quad \delta\!K(n)\,=\,
K(n)\,-\,K(n+1),
\]
\begin{equation}\label{af}
\begin{array}{lll}
p(l)\,=\,\eta^l\,\, K(l+1)\;, &  \quad \mathrm{for }{\ }  l\geq 1\;, &  \quad p(0)=0\;, \\
q(l)\,=\,(1-\eta)^l\,\, K(1)^{l-1}\;,  & \quad \mathrm{for } {\ } l\geq 1\;, & \quad q(0)=0\;, \\
r(l)\,=\,\eta^l\,\left[\eta\,\,  \delta\!K\,(l+1)\,+\, (1-\eta)\,
K(l+1)\,\, \delta\!K(0)\right]\;, & \quad \mathrm{for } {\ } l\geq
1, & \quad
r(0)=0\;,\\
\rho=\eta \,\, \delta\!K(1)+ (1-\eta)\,\, K(1)\,\, \delta\!K(0)\;.
& &
\end{array}
\end{equation}
Then we find
\begin{equation}\label{PS}
 \hat{P}(s)= \delta\!K(0)+\rho s+ \frac
 {s}{1-\hat{p}(s)\hat{q}(s)}\left[
\hat{r}(s)+\rho\hat{p}(s)\hat{q}(s)+\rho\,\,
K(1)\hat{q}(s)+K(1)\,\, \hat{q}(s)\hat{r}(s)
 \right]\;,
\end{equation}
where $\hat{p}(s),\hat{q}(s),\hat{r}(s)$ are the generating
functions corresponding to $p(l), q(l), r(l),$ respectively.

In the marginal cases $\eta=0$ and $\eta=1$, the probability
$P(\tau)$ can be readily calculated,
\begin{equation}\label{Peta}
 P_{\eta=0}(\tau) \,=\, K(1)^{\tau}\,\,  \delta\!K(0), \quad
 P_{\eta=1}(\tau) = \delta\!K(\tau).
\end{equation}
The above equation shows that in the case of $\eta\,=\,0,$ for
\textit{any choice} of the pdf $F$ and $Q$, the probability
$P(\tau)$ decays exponentially. In the opposite case $\eta\,=\,1,$
many different types of behavior are possible, depending upon the
particular choice of $F$ and $Q$.

To estimate the upper and lower bounds for $P(\tau)$ for any
$\eta$, one can use the fact that
$$
K(1)^n \le K(n) \le K(1) \;\;\;\; \mbox{and}\;\;\;\; 0 \le
\delta\!K(n) \le K(1)\;, \;\;\;\;\; n\,\in\, \mathbb{N}.
$$
Then the upper bound for $K(n)$ is trivial, since $0 \le F(u) \le
1$ for any $u \in [0,1]$. The upper bound for $K(n)$ exists if the
interval of the random variable $u$ is bounded and therefore can
be mapped onto $[0,1]$ (as a consequence of Jensen's inequality,
and of the fact that the function $u:\rightarrow u^n$ is convex on
the interval $]0,1[$ for any integer $n$). The calculation given
in \cite{FVL} allows for the following estimation for the upper
bound,
\begin{eqnarray} \label{Pupper}
P_{\eta}(\tau) &\le& \eta^{\tau} \,\, \delta\!K(\tau)
    + (1-\eta) K(1)\,\, \delta\!K(0)
    \left[ \eta + (1-\eta)\,\,  K(1) \right]^{\tau-1} \\ \nonumber
    &+& \eta \,\, K(1) \,\, \left\{ \left[ \eta + (1-\eta)\,\,  K(1) \right]^{\tau-1}
    - \eta^{\tau-1} \right\} \;,
\end{eqnarray}
and, for the lower bound,
\begin{eqnarray} \label{Plower}
P_{\eta}(\tau) &\ge& \eta^{\tau}\,\,  \delta\!K(\tau) + (1-\eta)\,\,  K(1)^{\tau} \,\, \delta\!K(0) \\
\nonumber
    &=& \eta^{\tau}  P_{\eta=1}(\tau) + (1-\eta) P_{\eta=0}(\tau)\;.
\end{eqnarray}
We thus see that, for any $0 \le \eta < 1$, the decay of
distribution $P(\tau)$ is bounded by exponentials. Furthermore,
the bounds (\ref{Pupper}) and (\ref{Plower}) turns into exact
equalities, in the marginal cases $\eta = 0$ and $\eta = 1$.

\begin{figure}[ht]
\noindent
\begin{minipage}[b]{.36\linewidth}
\begin{center}
\epsfig{file=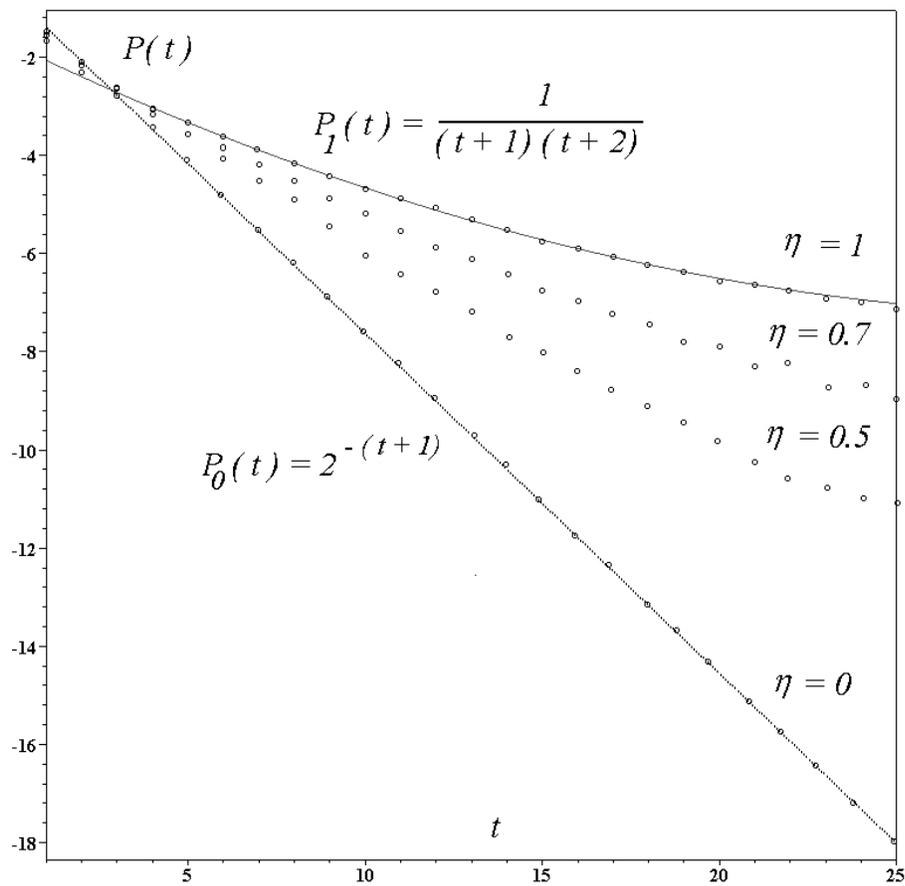, angle= 0,width=12.0cm}
\end{center}
\end{minipage}
\caption{\label{fig6}. The distributions of wandering times near
the wall in the discrete time model of Sec.~\ref{Sec6}, in the
case of the uniform densities $dF(u)\, = \,dG(u)\, =\, du$ for all
$u\in [0,\infty)$ at different values of control parameter
$\eta$.}
\end{figure}

The simpler and explicit expressions can be given for $P(\tau)$
provided the densities are uniform  $dF(u) = dQ(u) = du$ for all
$u\in [0,1]$. Then the equations (\ref{Peta}) give,
\begin{equation}\label{Ptunif}
  P_{\eta=0}(\,\tau\,)\,=\, 2^{-\,(\tau\,+\,1)}, \quad
P_{\eta=1}(\,\tau\,)\,=\, \frac 1{(\tau\,+\,1)(\tau\,+\,2)}.
\end{equation}
For the intermediate values of $\eta$, the upper and lower bounds
are
\begin{equation}\label{Ptbounds}
\frac{\eta^{\tau}}{(\tau\,+\,1)(\tau\,+\,2)} \, + \,(1-\eta)\,
2^{-\,(\tau\,+\,1)} \,\le \, P(\,\tau\,)\le \frac 12 \,\,
\left(\,\frac{1\,+\,\eta}{2}\,\right)^{\tau}.
\end{equation}
The above results are displayed in Fig.~\ref{fig6}.

\section{Conclusion }
\label{Sec8}
\vspace*{-0.5pt}
\noindent

In the present paper, we have considered the stochastic model of
turbulent transport in the SOL of tokamak. This problem allows for
the concurrent symmetries, and the system exhibits a sever
instability with respect to both perturbations either particle
density or vorticity. Instability would reveal itself in the
appearance of high density blobs of particles hitting into the
reactor wall.

The accounting for the dissipation processes introduces the order
parameter $\xi = \left|k_y\right|\left/ \left(k^2_x+k_y^2\right)
\right.$ and its critical value $\xi_c$ such that the particle
density fluctuation $\delta\!n (\xi)$ grows unboundedly with time
as $\xi\,>\,\xi_c$ and damps out otherwise. In the present paper,
we compute the value of $\xi_c,$ in the first order of
perturbation theory developed with respect to the small parameter
$\left.\rho_s\,\right/ R \bar{n}$ where $\rho_s$ is the Larmor
radius, $R$ is the major radius of torus, and $\bar{n}$ is the
mean normalized density of particles.

Our results demonstrate convincingly that the possible
correlations between density and vorticity fluctuations would
drastically change the value $\xi_c$ modifying the stability of
model. Characterizing the possible reciprocal correlations between
the density and vorticity fluctuations by the specific correlation
time $\tau_c$, we demonstrate that any fluctuation of particle
density  grows up with time in the large scale limit ($k\to 0$) as
$\tau_c \to \infty$ (the density and vorticity fluctuations are
uncorrelated) and therefore $\xi_c = 0.$ Alternatively, $\xi_c
>0$  provided $\tau_c<\infty$.

The reciprocal correlations between the fluctuations in the
divertor is of vital importance for a possibility to stabilize the
turbulent cross field system, in the large scales, by  biasing the
limiter surface discussed in the literature before
\cite{Ghendrih}. Namely, if $\xi_c > 0$, there would be a number
of intervals $[V_{k-1},V_k]$ for the uniform electric poloidal
drifts $V$ such that all fluctuations arisen in the system are
damped out fast. In particular, in the first order of perturbation
theory, there exists one threshold value $V_c$ such that the
instability in the system is bent down as $V\,>\,V_c.$ However,
$V_c \to \infty$ as $\xi_c\to 0.$

To get an insight into the statistics of growing fluctuations of
particle density that appear as high-density blobs of particles
close to the reactor wall, we note that their growth rates are
determined by the advanced Green's functions analytical in the
lower half plain of the frequency space. We replace the rather
complicated dynamical process of creation and convection of
growing density fluctuations  by the turbulent flow with the
problem of discrete time random walks concluding at a boundary.
Such a substitution can be naturally interpreted as a Monte Carlo
simulation procedure for the particle flux. Herewith, the
wandering time spectra which determine the pdf of the particle
flux in such a toy model are either exponential or bounded by the
exponential from above. This observation is in a qualitative
agreement with the numerical data reported in \cite{Ghendrih}.

\section{Acknowledgment}
\noindent

Authors are grateful to the participants of the Journ\'{e}es de
Dynamique Non Lin\'{e}aire on  2 December 2003, Luminy, Marseille
(France) for the valuable discussions. One of the authors (D.V.)
deeply thanks the Centre de Physique Theorique (CNRS), Marseille,
where the present work had been started.

\centerline{\textbf{Table 1: Canonical dimensions of fields and
parameters in the action functional (\ref{S})}}
\begin{center}
\begin{tabular}{l c c c c c c c c c c c c }
\hline \hline
   & $\nu$ & $u_0$  & $n$  & $w$ & $\mathbf{v}$&$\mathbf{v'}$&$n'$&$w'$&$\phi$&$g_k$\\
   \hline
     $d^k$ & -2 & 0 & 0 &0  & -1 & $d$ & $d$ & $d$& -2  & $-1 $ \\
  $d^\omega$ & 1 & 0 &0  &1  & 1  & 0 & 0 & 2& 1 &2\\
  $d$ & 0 & 0 & 0 & 2 & 1 & $d$ & $d$ & $d+4$ & 0 &3 \\
  \hline
  \hline
\label{Tab1}
\end{tabular}
\end{center}


\section{Figures:}
\noindent

\end{document}